\documentclass[a4paper,11pt]{article}

\usepackage{pos}
\usepackage{siunitx}
\usepackage{xspace}
\usepackage{multicol}

\newcommand{\eV}[1]{\ensuremath{10^{#1}\,\text{eV}}\xspace}

\newcommand{\numu}{\ensuremath{\nu_\mu}\xspace}
\newcommand{\numubar}{\ensuremath{\overline{\nu}_\mu}\xspace}

\newcommand{\Edep}{\ensuremath{E_\mathrm{deposit}}\xspace}

\title{Estimating the coincidence rate between the optical and radio array of IceCube-Gen2}
\ShortTitle{Coincidence events with IceCube-Gen2}

\author{The IceCube-Gen2 Collaboration \\{\normalsize \normalfont(a complete list of authors can be found at the end of the proceedings)}\\}

\emailAdd{felix.schluter@icecube.wisc.edu}
\emailAdd{simona.toscano@icecube.wisc.edu}

\abstract{The IceCube-Gen2 Neutrino Observatory is proposed to extend the all-flavour energy range of IceCube beyond PeV energies. It will comprise two key components: I) An enlarged 8$\,$km$^3$ in-ice optical Cherenkov array to measure the continuation of the IceCube astrophysical neutrino flux and improve IceCube's point source sensitivity above $\sim\,$100$\,$TeV; and II) A very large in-ice radio array with a surface area of about 500$\,$km$^2$. Radio waves propagate through ice with a kilometer-long attenuation length, hence a sparse radio array allows us to instrument a huge volume of ice to achieve a sufficient sensitivity to detect neutrinos with energies above tens of PeV.

The different signal topologies for neutrino-induced events measured by the optical and in-ice radio detector - the radio detector is mostly sensitive to the cascades produced in the neutrino interaction, while the optical detector can detect long-ranging muon and tau leptons with high accuracy -  yield highly complementary information. When detected in coincidence, these signals will allow us to reconstruct the neutrino energy and arrival direction with high fidelity. Furthermore, if events are detected in coincidence with a sufficient rate, they resemble the unique opportunity to study systematic uncertainties and to cross-calibrate both detector components. 

We present the expected rate of coincidence events for 10 years of operation. Furthermore, we analyzed possible detector optimizations to increase the coincidence rate.

\vspace{4mm}
{\bfseries Corresponding authors:}
Felix Schlüter$^{1*}$, Simona Toscano$^1$\\
{$^{1}$ \itshape Inter-University Institute For High Energies (IIHE), Université libre de Bruxelles (ULB) \\
  Boulevard du Triomphe 2, 1050 Brussels, Belgium}\\
  $^*$ Presenter

\ConferenceLogo{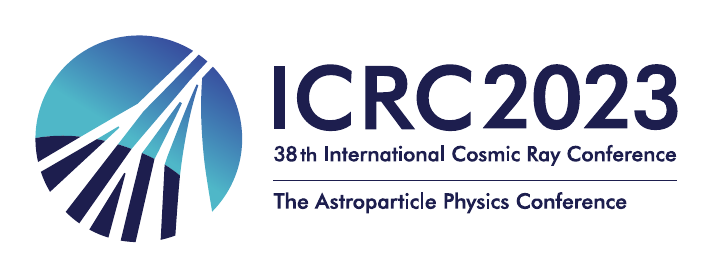}

\FullConference{The 38th International Cosmic Ray Conference (ICRC2023)\\ 26 July -- 3 August, 2023\\ Nagoya, Japan}

}

\begin{document}
\maketitle


\section{Introduction}

Recently, the IceCube Neutrino Observatory found the first direct evidence of TeV neutrino emission from a nearby active galaxy \cite{IceCube:2022der}. However, to identify the sources of ultra-high-energy cosmic rays ($E \ge 10^{18}\,$eV) and study particle acceleration and interactions at those energies, neutrinos with energies $E \ge 10\,$PeV have to be detected. The IceCube-Gen2 facility aims to enhance the observatory's sensitivity to neutrinos at such energies \cite{IceCube-Gen2:2020qha}. It will comprise 160 additional strings with optical modules spaced by \SI{240}{m} with a total instrumented volume of \SI{8}{km^3} and an array of several hundred, sparsely spaced radio stations covering a surface area of  $\sim$\SI{500}{km^2}.

The radio detector is sensitive to the nanosecond-long broadband radio pulses produced by particle cascades induced by all neutrino interactions as well as by catastrophic energy losses of muon and tau leptons produced in charge-current interactions during the propagation \cite{Garcia-Fernandez:2020dhb}. To produce a radio signal above the thermal background, energies above $\gtrsim \SIrange{1}{10}{PeV}$ are necessary. The radio emission does not experience any significant scatter and has an attenuation length on the order of one kilometer. This allows a single radio-antenna station to probe a volume of several cubic kilometers of ice. In contrast to that, the optical detector is only sensitive to the energy deposit of particles contained in or passing through the instrumented detector volume. 

Due to the different signal topologies between the two detection techniques, coincidence measurements of neutrino interactions with the radio and optical detector yield highly complementary information which improves the reconstruction of the neutrino properties, in particular their energy. For example, when a muon neutrino interacts outside the optical detector and then passes through it, it is possible to determine the neutrino arrival direction to sub-degree accuracy while the neutrino energy can only be estimated with large uncertainty with the optical array alone. With the radio detector it will be possible to measure interaction vertex and energy transferred into the hadronic cascade \cite{RNO-G:2020rmc}, which will significantly improve the energy reconstruct for primary neutrino. 
Besides that, coincidence detections of neutrinos between the radio and optical detector could be to cross-calibrate the two detectors and lower systematic uncertainties. 
In this work, we investigate the potential of coincidence detections of ultra-high-energy (UHE) neutrinos with the radio and optical detector of IceCube-Gen2. 


\begin{figure}[tbp]
    \centering
    \includegraphics[width=0.9\linewidth]{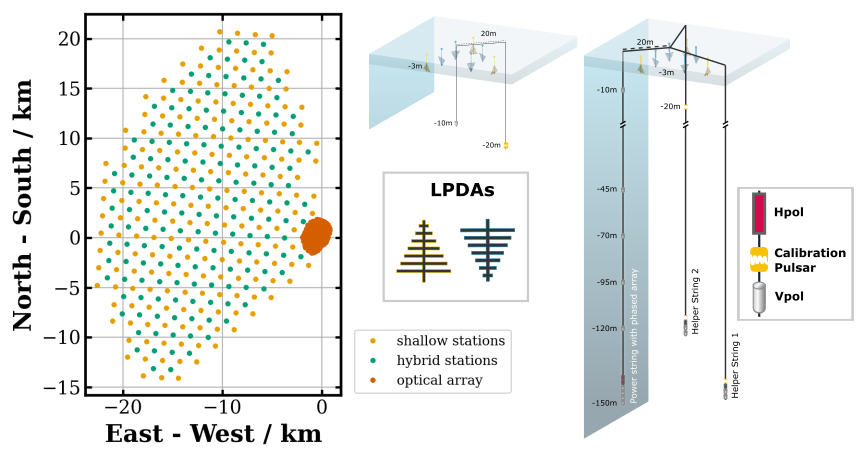}
    \caption{Preliminary ``baseline'' array layout (\textit{bottom-left}) and station designs (\textit{top-left}: shallow station, \textit{right}: hybrid station) of the IceCube-Gen2 radio detector.
    }
    \label{fig:array_layout}
\end{figure}
\section{The IceCube-Gen2 radio detector}
\label{sec:detector}
The design of the IceCube-Gen2 radio detector is informed by the pilot experiments ARIANNA \cite{ARIANNA:2019scz} and ARA \cite{ARA:2015wxq}, which pioneered slightly different detector concepts, as well as RNO-G \cite{RNO-G:2020rmc}, currently constructed in Greenland, which combines both concepts. ARIANNA uses ``shallow'' antennas deployed a few meters beneath the surface, while ARA deployed ``deep'' antennas in \SI{200}{m} deep boreholes. The Gen2 radio detector will utilize shallow and deep antennas in two different station designs: A shallow station which mostly utilized high-gain LPDA antennas and a hybrid station that combines the shallow antennas with deep antennas (similar to RNO-G). The deep antennas are comprised of low-gain vertical and horizontally polarized dipole antennas (Vpols \& Hpols) which can fit in narrow boreholes that reach \SI{150}{m} deep. Both station types comprise calibration pulsers and will use the same read-out electronics which is sensitive between about \SIrange{80}{620}{MHz}. A schematic of both station types can be found in Fig.~\ref{fig:array_layout}. Several array layouts with varying numbers of shallow and hybrid stations and different spacings between those are currently under investigation. Fig.~\ref{fig:array_layout} shows the ``baseline'' layout which features 164 hybrid stations with a spacing of \SI{1.75}{km} and 197 shallow stations with a spacing of \SI{1.24}{km}.
The hybrid stations feature a 4-channel phased array on the bottom of the ``power string'' (cf.\ Fig.~\ref{fig:array_layout}). The signals of the 4 Vpol antennas are constantly phased in several beams sensitive to different directions in zenith. This will increase the signal-to-noise ratio for correlated signals arriving from those directions at the antennas while suppressing thermal noise fluctuations. This concept has been developed and tested by ARA \cite{Allison:2018ynt}.


\section{Simulating radio-optical coincidence events for IceCube-Gen2}
\label{sec:simulations}

We simulate the neutrino interactions in the ice, the radio emission produced by those interactions, and the instrumental response of the radio detector with the open-source framework NuRadioMC \cite{Glaser:2019cws}. For the set of neutrino events that triggered the radio detector,
we simulate the optical ``counterpart'' using IceCube's simulation and reconstruction framework IceTray \cite{DeYoung:2005zz}. 

\textbf{Neutrino generation:}
We simulate neutrinos of all flavors with discrete energies between \eV{16} and \eV{20} in bins of $\log_{10}(\Delta E/\mathrm{eV}) = 0.5$ undergoing charge-current (CC) and neutral-current (NC) interactions\footnote{The ratio between CC and NC interactions is $\sim 70\%$ to 30\% which in agreement with their cross-sections.}. The arrival directions are isotopically distributed , and the neutrino interaction vertices are randomly distributed within a rectangular volume. The height of the rectangle is \SI{2800}{m}, i.e., roughly the thickness of the polar ice sheet. The area depends on the energy, flavor, and interaction type of the event: For all cascade-like events (those not producing a muon or tau lepton in the first interaction), the considered area is \SI{34}{km} along the East-West direction and \SI{50}{km} along the North-South direction. For track-like events (those with a muon or tau lepton produced in the first interaction), the area is dynamically increased depending on the zenith angle and energy to allow for all possible vertices from which the track-like leptons could reach the original rectangular volume defined for the cascade-like events. 
All produced track-like leptons are propagated using the PROPOSAL code \cite{Koehne:2013gpa,Dunsch:2018nsc} and their energy losses above \SI{1}{PeV}, i.e., those relevant for the emission of radio signals, are stored.
Each neutrino event is assigned a weight $w_i$ which describes the neutrino's probability to arrive at the vertex location considering its trajectory through the Earth and/or ice sheet\footnote{This weight does not include the probability of the neutrino to interact but only to have reached the interaction vertex.}. 
In total, we generated $3.4 \cdot 10^8$ neutrino interactions.

\begin{figure}[tbp]
    \centering
    \includegraphics[width=0.565\linewidth]{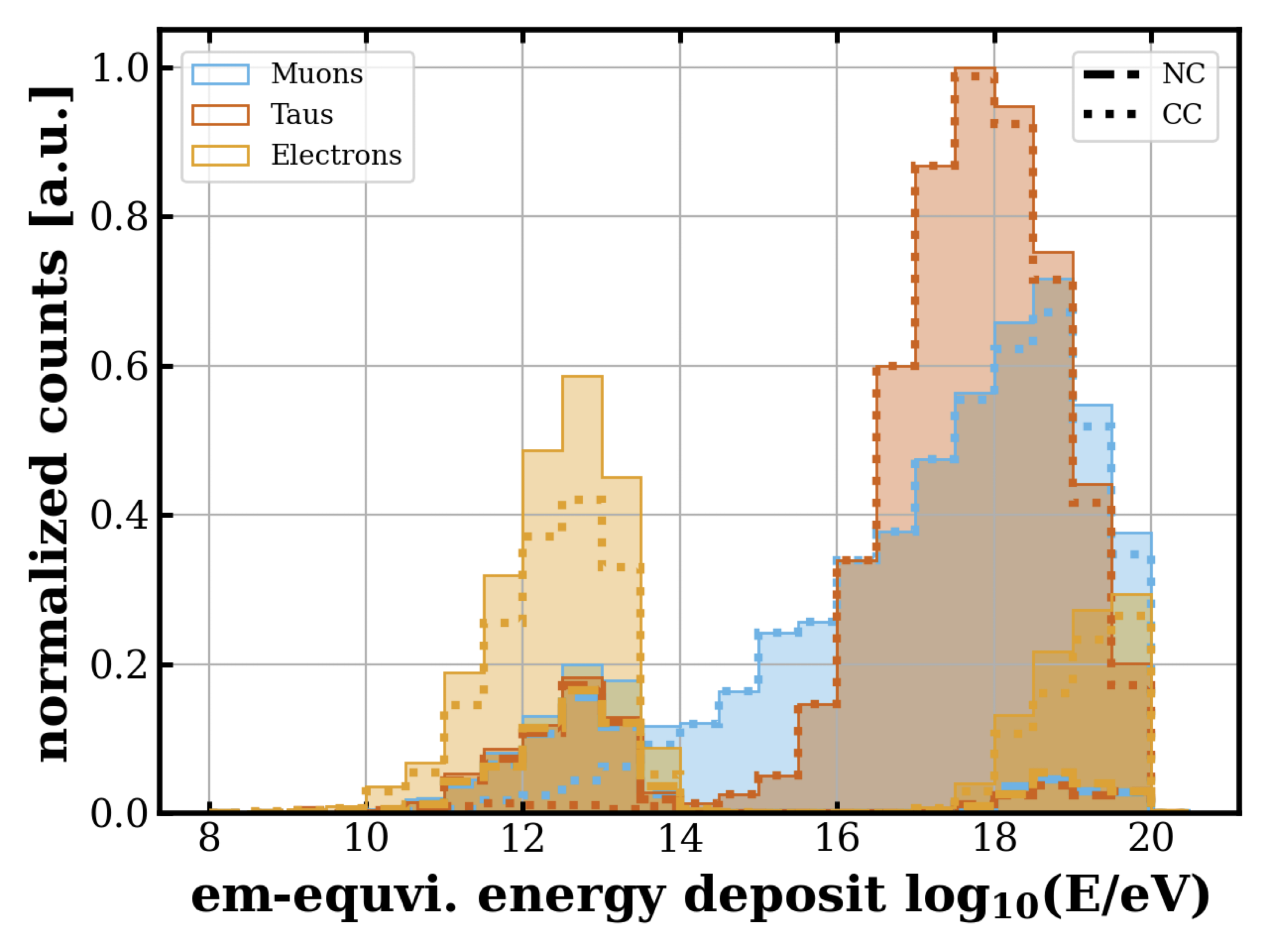}\hfill
    \includegraphics[width=0.42\linewidth]{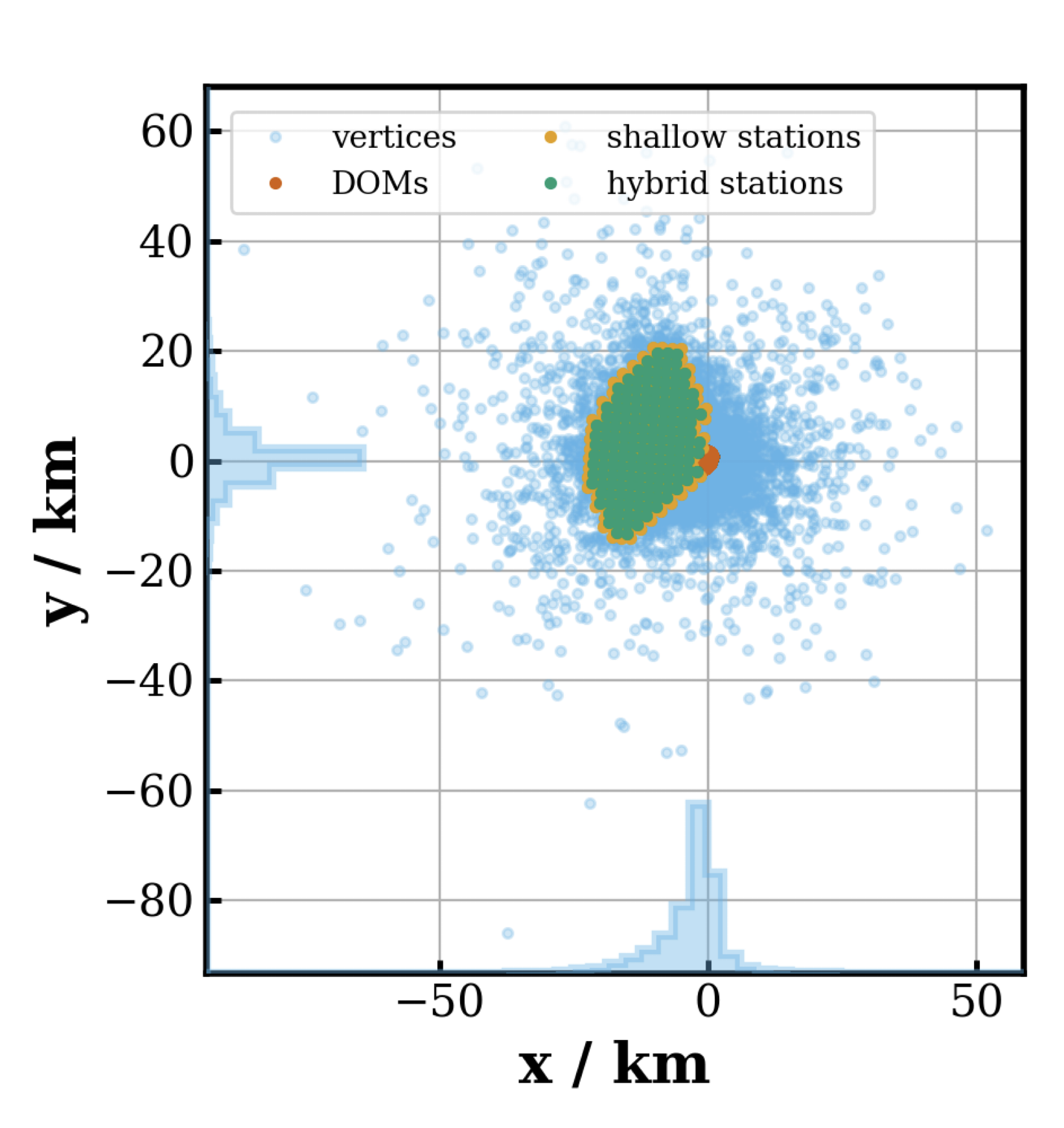}

    \caption{\textit{Left}: Normalized distribution of the electromagnetic-equivalent energy deposit. As in the following ``Muons'' in the label refers to both \numu and \numubar, the same holds for the other flavors. \textit{Right}: Unweighted scatter plot of the neutrino interaction vertices for all coincidence events.}
    \label{fig:energy_deposit}
\end{figure}

\textbf{Radio simulation:}
For each shower produced by a neutrino interaction or lepton, the electric field vector at every antenna\footnote{To increase computation time we limit the simulated signals to antennas that are used for triggering.} within an energy-dependent distance around the shower is calculated. First, all possible trajectories of the radio emission between the shower and antenna are determined taking into account refraction and reflection. Then the radio emission produced by the shower at the antenna location is calculated using a semi-analytic model based on realistic charge-excess profiles of in-ice particle cascades \cite{Alvarez-Muniz:2010wjm}. Finally, the radio emission received at the antenna is propagated through the instrumental response of the detector and a trigger is simulated. We simulated two independent triggers: 1) a 2 out of 4 coincidence amplitude threshold trigger with the 4 downward facing LPDAs, and 2) a power-integrating threshold trigger on the beam-formed signals of the 4-channel phased array. The trigger thresholds are defined by a maximum trigger rate of \SI{100}{Hz} each, simulating thermal noise with a temperature of \SI{300}{K}. For a general and comprehensive description of the entire procedure and all steps involved, see \cite{Glaser:2019cws}. In total, $1.1 \cdot 10^6$ events triggered the radio detector.

\textbf{Optical simulations:}
To simulate triggers of the optical detector, first, we determine if 1) a cascade-like event is produced within a ``fiducial'' volume of \SI{150}{m} around the optical array, or 2) if a track-like lepton intersects with the same fiducial volume.
And second, we determined the electromagnetic-equivalent energy deposited \Edep, which is proportional to the produced Cherenkov light \cite{Radel:2012ij} inside the fiducial volume\footnote{For this calculation we do not take into account the spatial expansion of the cascades.} . Figure \ref{fig:energy_deposit} (\textit{left}) shows the normalized distribution of \Edep separated by flavor (color-coded) and interaction type. The normalization reflects expected event rate of coincidence events per flavor for the simulated, unrealistically hard energy spectrum.
For each lepton flavor, the deposited energy inside the fiducial volume shows two distinct peaks: one at TeV energies and one at EeV energies. At high energies, the events are comprised of mostly muon or tau leptons originating from CC-interactions, that intersect and deposit their energy inside the fiducial volume. A second component is found from electron-neutrino-induced showers contained inside the fiducial volume. At lower energies, the observed energy deposits mostly originate from low-energy secondary muons produced in hadronic cascades (all flavor NC + electron CC events) that are found outside the detector. Those secondary muons carry only a small fraction ($\sim 10^{-6}$) of the initial neutrino energy, and hence, although they originate from the cascades induced by the highest energy neutrinos, the deposited energy inside the fiducial volume is found at much lower energies. To determine a trigger condition for the optical detector based on the deposited energy, we assume that the sensitivity of the Gen2 optical array is similar to that of the current IceCube Gen1 array. Although the spacing between strings is doubled in Gen2, the efficiency of the Gen2 optical modules is expected to improve by a factor of $\sim 3$ w.r.t. the Gen1 modules, compensating for the larger spacing. As such, as a conservative criterion, we require $\Edep > 100\,$TeV to consider an event as detectable by the Gen2 optical array. This condition rejects around 25\% of all previously selected events which are mostly made up of the low-energy secondary muons\footnote{The simulation of muons from hadronic cascades uses an energy-independent parameterization, obtained from CORSIKA simulations at lower energies, which is scaled linearly with energy. This procedure likely underestimates the number of muons at energies relevant is this work.}. 
Due to the limitations of the hybrid simulations, the leptons from the initial neutrino interaction are propagated twice: First in NuRadioMC to simulated (catastrophic) energy losses with energies above \SI{1}{PeV} relevant for the radio detector, and again in IceTray to determine if the events trigger the optical detector. The repeated simulations will be different on an event-by-event level\footnote{Also, the PROPOSAL version within IceTray is different w.r.t. NuRadioMC.}, which could potentially bias our results: leptons with stronger energy losses (before entering the optical detector) are more likely to trigger the radio detector but less likely to trigger the optical detector. However, we found no indication for such bias in our results, i.e., in the considered energy range leptons which undergo strong energy losses will still have enough energy to reach the optical detector. Note, that this effect would only negatively affect the coincidence rate.
\begin{figure}[tbp]
    \centering
    \includegraphics[width=0.49\linewidth]{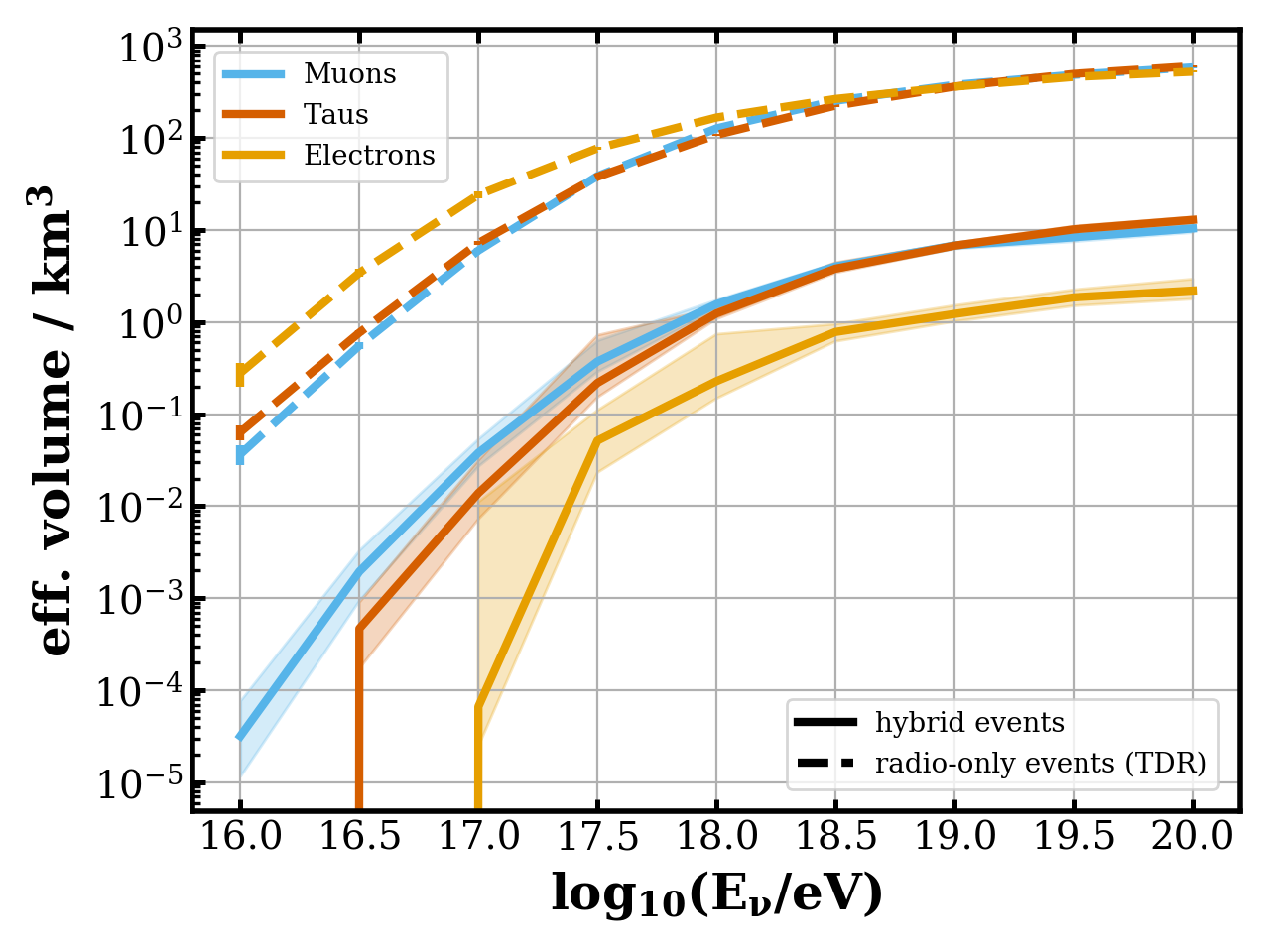}\hfill
    \includegraphics[width=0.49\linewidth]{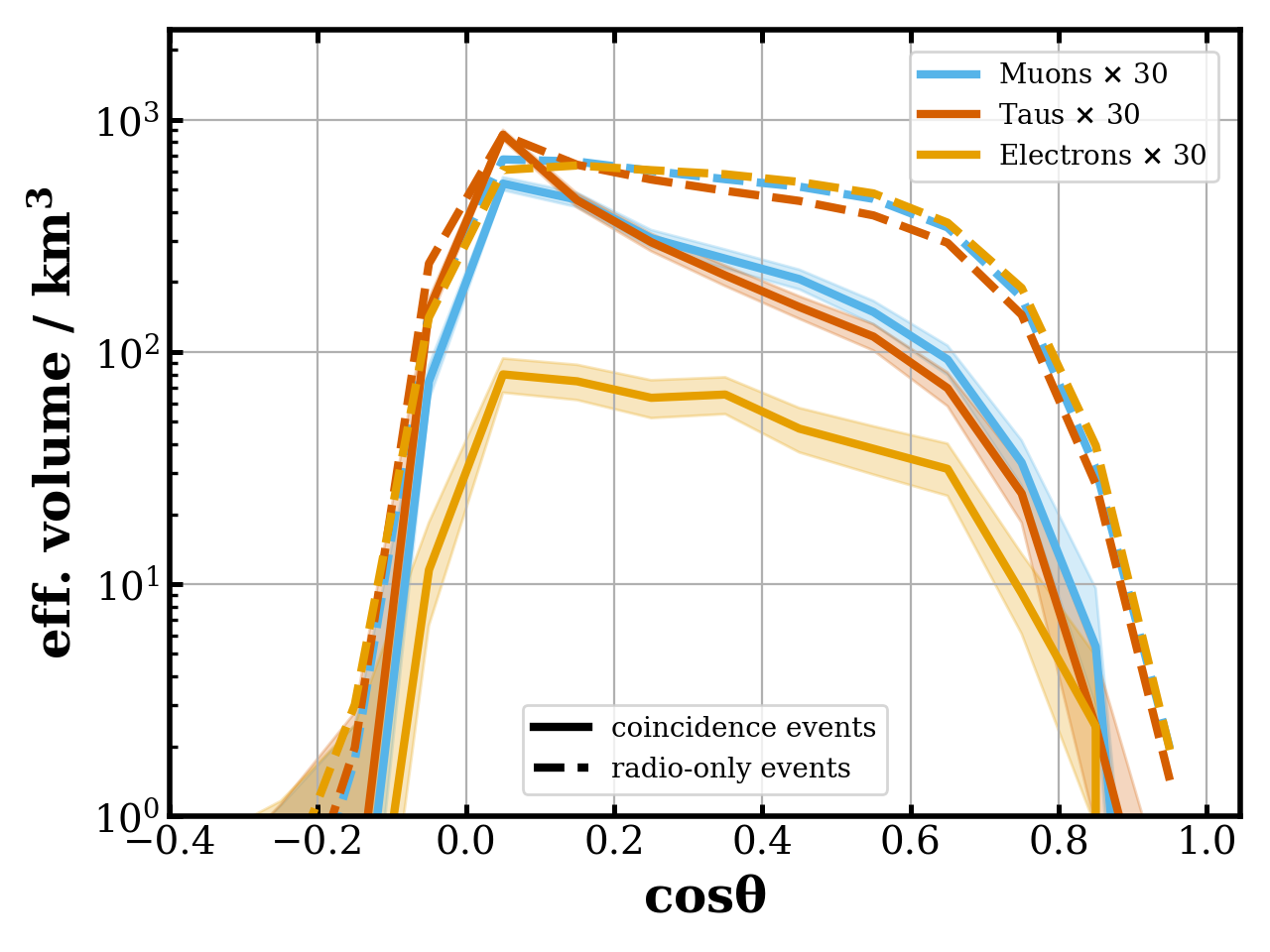}
    \caption{Per flavor effective volume for coincidence events (solid lines) and radio-only events (dashed lines) as a function of the energy (\textit{left}) and zenith angle (\textit{right}). Colored band represent statistical uncertainties. {\it Right}: Effective volume for coincidence events is multiplied by 30 to allow a better comparison.}
    \label{fig:eff_volume}
\end{figure}
\section{Rate of coincidence events}
\label{sec:analysis}
\begin{figure}[tbp]
    \centering
    \includegraphics[width=0.49\linewidth]{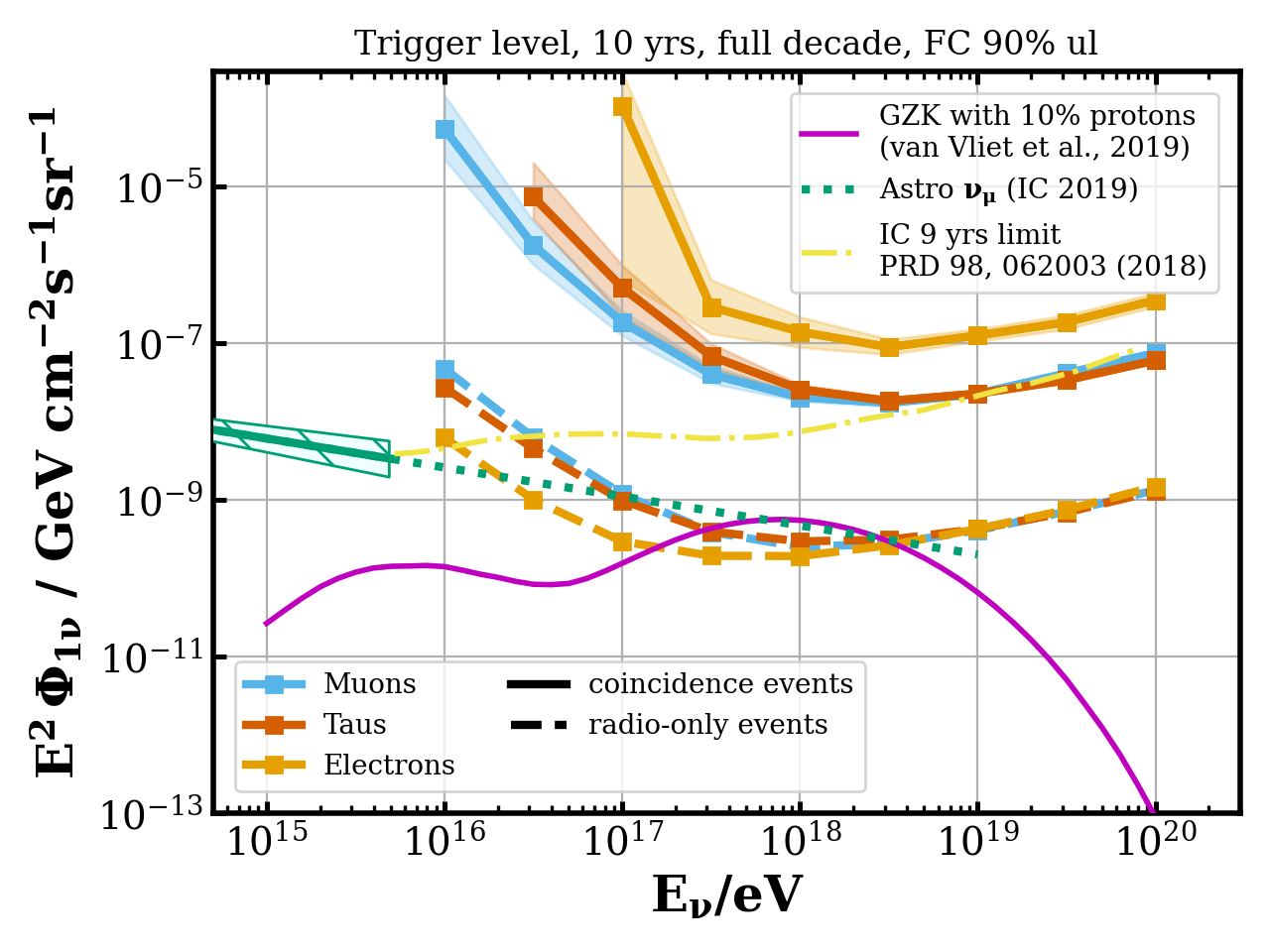}\hfill
    \includegraphics[width=0.49\linewidth]{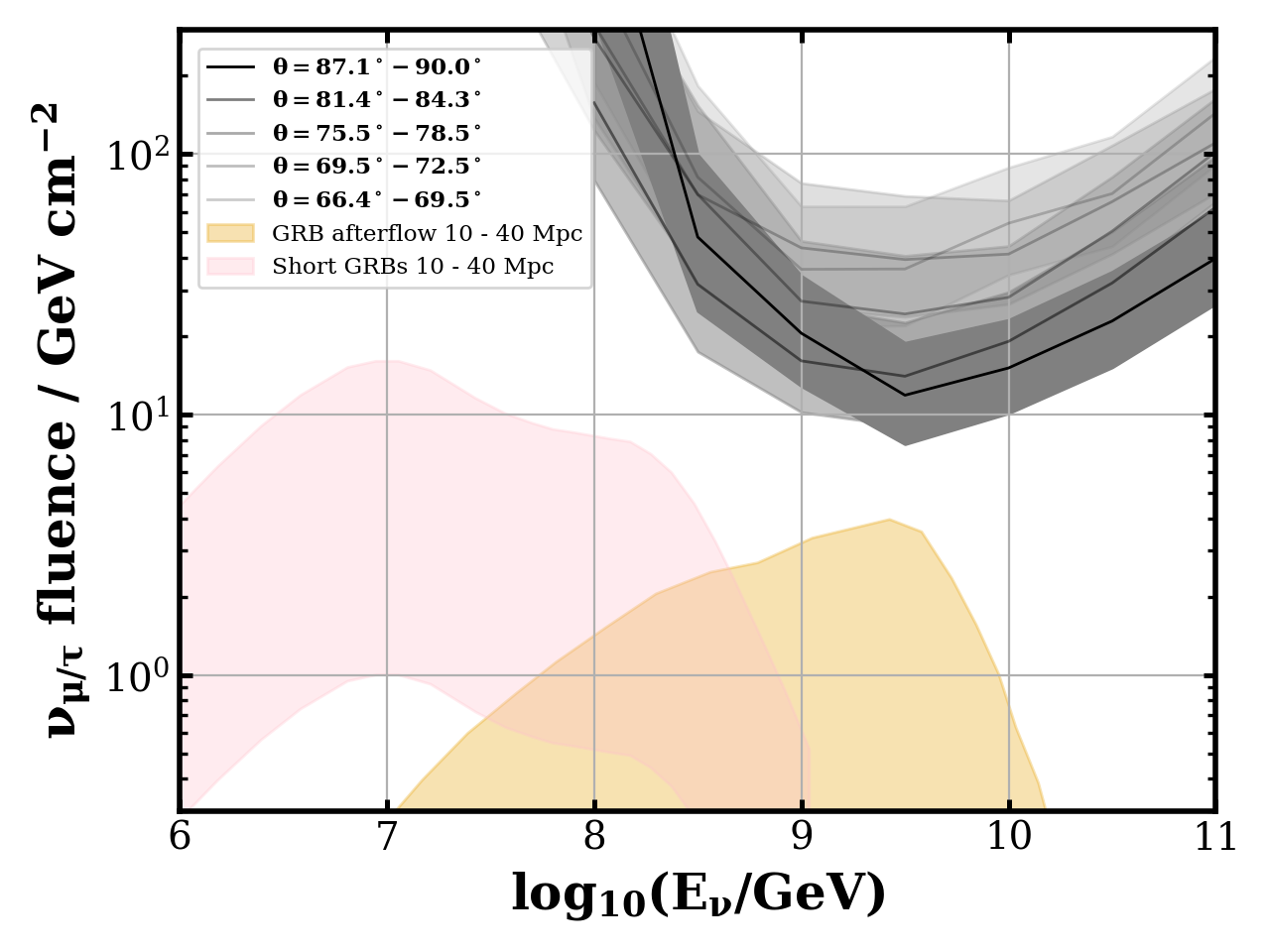}

    \caption{\textit{Left}: Single flavor sensitivity to a diffuse emission for coincidence (dashed lines) and radio-only (solid lines) events. The flux limit is compared to different flux predictions. \textit{Right}: Expected instantaneous sensitivity for coincidence events and neutrino fluence ($\nu_\mu + \nu_\tau$) for two different source classes.}
    \label{fig:sensitivity}
\end{figure}

In the following, we analyse 14257 selected coincidence events. Figure \ref{fig:energy_deposit} (\textit{right}) shows a map of all neutrino-interaction vertices together with the positions of the radio stations and optical modules. The $x-$ and $y-$axis show a histogram of the projected, unweighted distributions. The majority of coincidence neutrinos have their interaction relatively close to the optical array. Considering only events with a trigger from the radio hybrid station closest to the optical array accounts for roughly $1 / 3$ of all coincidence events. We can also identify a few events where the interaction vertex is tens of kilometers away from any radio station. Those are track-like events that have triggered the radio detector from a catastrophic energy loss of the charged lepton.
With the selected coincidence events we determine the effective volume $V_\mathrm{eff}$ which depends on the energy $E$ and zenith angle $\theta$ of the primary neutrinos
\begin{equation}
    V_\mathrm{eff}(E, \theta) = V_\mathrm{gen}(E, \theta) \frac{\sum_i^{n_\mathrm{coinc}(E, \theta)} w_i}{n_\mathrm{gen}(E, \theta)}.
    \label{eq:veff}
\end{equation}
The index ``gen'' refers to the number of generated neutrino interactions and the volume in which they are generated. 
The energy bins are that of the simulated neutrinos, the zenith angle bins are equidistant in $\cos \theta$ with a width of 0.1. The effective volume per flavor as a function of the energy averaged over all arrival directions is shown in Fig.~\ref{fig:eff_volume} (\textit{left}). The solid lines indicate the here-derived effective volume for coincidence events while the dashed lines show the effective volume for all triggered radio-only events from \cite{IceCubeGen2TDR}. The error bands show the 68\% statistical uncertainty according to the Feldman-Cousins method \cite{Feldman:1997qc}. At lower energies, muon neutrinos provide the highest sensitivity because the muons propagate the furthest through the ice while taus decay after a while. At higher energies, taus exceed muons in range. Since electron neutrinos have to interact inside the optical detector their effective volume is much reduced, and there is a higher energy threshold for the detection because the radio emission has to reach the nearest radio antennas from the deep optical array. On the right side of Figure \ref{fig:eff_volume} the effective volume as a function of the arrival direction, i.e., averaged over all energies, is shown. The effective volume for coincidence events is scaled by a constant arbitrary factor to allow a better comparison with the result for radio-only events. For both event sets, the highest sensitivity is just above the horizon. Below the horizon, the survival probability for neutrinos passing through the Earth limits the sensitivity. For vertical arrival directions the lower target mass in front of the antennas limits the sensitivity. For coincidence events, the additional geometrical constrain that the leptons have to intersect with the optical detector volume reduces the sensitivity for vertical events further. With the estimated effective volume, we can now derive a flux sensitivity limit $\Phi_\mathrm{UL}(E)$ for a given live time $t_\mathrm{live}$ and energy range $\Delta\left(\log_{10}E\right)$ 
\begin{equation}
    \Phi_\mathrm{UL}(E) = \frac{2.44}{4 \pi \, [V_\mathrm{eff}(E) / L_\mathrm{int}(E)] \, t_\mathrm{live} \, \log(10) \, \Delta\left(\log_{10}E\right) \, E}.
    \label{eq:flux}
\end{equation} 
The interaction length of neutrinos $L_\mathrm{int}(E)$ in ice is determined using the parameterization of the cross-section from \cite{Connolly:2011vc}. The 2.44 counts reflect the 90\% confidence interval after the Feldman-Cousins method. The term $ 4 \pi V_\mathrm{eff}(E) / L_\mathrm{int}(E)$ can be interpreted as the effective area which is commonly used by other neutrino telescopes. Figure \ref{fig:sensitivity} (\textit{left}) shows the sensitivity per flavor, 10 years of operation, and a full decade in energy for radio-only and coincidence events. Those limits are compared to the flux detected by IceCube which is extrapolated to higher energies \cite{IceCube:2021uhz}, as well as a model for cosmogenic neutrinos \cite{vanVliet:2019nse} and the limit obtained from the IceCube extremely-high-energy neutrino search using 9 years of data \cite{IceCube:2018fhm}. For the two flux models, we expect 0.2 neutrinos each integrated over all energies and flavors. Hence, the sensitivity is most likely not sufficient to observe coincidence events from a diffuse emission within a reasonable time. We also studied the sensitivity toward transient event classes. In Figure \ref{fig:sensitivity} (\textit{right}) the prediction of the neutrino fluence from two different source classes for muon and tau neutrinos is shown together with the sensitivity of Gen2 for coincidence events in different zenith angle bands. The shaded bands describe the expected azimuthal asymmetry in the sensitivity which is around $\pm 25\%$ for muon and tau neutrinos with a maximum for neutrinos coming from the West. For the chosen source classes the sensitivity is likely too low to see neutrinos in coincidence.

\section{Conclusion \& Discussion}
\label{sec:discussion}
The sensitivity of IceCube-Gen2 for coincidence events between the radio and optical detector is likely too low to observe those events in the simulated configuration. Therefore, in the following we discuss how the sensitivity can be improved.
Just above the horizon, the contribution from radio triggers of the deep and shallow station components is about equal, while events which have radio triggers in both components dominate\footnote{Typically, those triggers are not from one hybrid station but from several stations.}. At more down-going geometries, radio triggers for the deep component dominate, while just below the horizon the shallow component triggers contribute more. Overall, for the chosen array layout the deep triggers have a higher contribution. We repeated this analysis with simulations of an alternative radio detector layout (``hex-shallow'' cf.\ \cite{IceCubeGen2TDR}) which features more stations with a higher fraction of shallow stations (123 hybrid stations on \SI{2}{km} hex.\ grid, 414 shallow stations on \SI{1}{km} hex.\ grid). We found no noteworthy difference in the results presented here. Installing antennas deeper as currently planned, i.e., at depths of 200 to 400 meters or even deeper \cite{Bishop:2021sst}, will improve the coincidence rate. However, to exactly quantify the effect we would need new simulations which is out of the scope of this analysis.
As mentioned earlier, the radio station (hybrid) closest to the optical detector has a large contribution to the overall sensitivity. It is easy to imagine that the current arrangement between the radio and optical array is not optimal. Ideally the optical detector would be central within the radio array. Due to constraints at the South Pole, this is not possible. However, to test the (all)most optimal scenario, we simulated such a configuration. The overall sensitivity improved by a factor of 2 - 3, a substantial but not sufficient improvement to detect a significant number of neutrinos.
As mentioned before, the applied trigger condition for the optical detector is conservative. It is likely, that energy threshold can be lowered considerably which increases the aperture for hadronic cascades outside of the optical detector as well as skimming track events. However, the accurate determination of the detection threshold demands simulations which comprise among other things a more precise modelling of hadronic cascades and photon propagation.
By increasing the sensitivity of the radio detector we expect to see, to first order, the same improvements in the coincidence rate. This could be accomplished in two ways: First, by increasing radio trigger rates which translate into the observation of weaker signals, for at least the closest radio station(s). Second, by developing smart algorithms which, at trigger level, can discriminate between thermal noise the dominant background and neutrino-induced signals. Currently, such an algorithm is developed for RNO-G \cite{Alan:2023}. However, even with a significant improvement of the coincidence sensitivity a detection would require an optimistic flux expectation of a cosmogenic neutrino beam.


\let\oldbibliography\thebibliography
\renewcommand{\thebibliography}[1]{%
  \oldbibliography{#1}%
  \setlength{\itemsep}{0pt}%
}

\begin{multicols}{2}
{\small
\bibliographystyle{ICRC}
\bibliography{references}
}
\end{multicols}

%

\clearpage

\section*{Full Author List: IceCube-Gen2 Collaboration}

\scriptsize
\noindent
R. Abbasi$^{17}$,
M. Ackermann$^{76}$,
J. Adams$^{22}$,
S. K. Agarwalla$^{47,\: 77}$,
J. A. Aguilar$^{12}$,
M. Ahlers$^{26}$,
J.M. Alameddine$^{27}$,
N. M. Amin$^{53}$,
K. Andeen$^{50}$,
G. Anton$^{30}$
% ...
\\
\\
% ...
$^{12}$ Universit{\'e} Libre de Bruxelles, Science Faculty CP230, B-1050 Brussels, Belgium \\
% ...
$^{17}$ Department of Physics, Loyola University Chicago, Chicago, IL 60660, USA \\
% ...
$^{22}$ Dept. of Physics and Astronomy, University of Canterbury, Private Bag 4800, Christchurch, New Zealand \\
% ...
$^{26}$ Niels Bohr Institute, University of Copenhagen, DK-2100 Copenhagen, Denmark \\
$^{27}$ Dept. of Physics, TU Dortmund University, D-44221 Dortmund, Germany \\
% ...
$^{30}$ Erlangen Centre for Astroparticle Physics, Friedrich-Alexander-Universit{\"a}t Erlangen-N{\"u}rnberg, D-91058 Erlangen, Germany \\
% ...
$^{47}$ Dept. of Physics and Wisconsin IceCube Particle Astrophysics Center, University of Wisconsin{\textendash}Madison, Madison, WI 53706, USA \\
% ...
$^{50}$ Department of Physics, Marquette University, Milwaukee, WI, 53201, USA \\
% ...
$^{53}$ Bartol Research Institute and Dept. of Physics and Astronomy, University of Delaware, Newark, DE 19716, USA \\
% ...
$^{76}$ Deutsches Elektronen-Synchrotron DESY, Platanenallee 6, 15738 Zeuthen, Germany  \\
$^{77}$ Institute of Physics, Sachivalaya Marg, Sainik School Post, Bhubaneswar 751005, India \\
% ...

\subsection*{Acknowledgements}

\noindent
The authors gratefully acknowledge the support from the following agencies and institutions:
USA {\textendash} U.S. National Science Foundation-Office of Polar Programs,
U.S. National Science Foundation-Physics Division,
U.S. National Science Foundation-EPSCoR,
Wisconsin Alumni Research Foundation,
Center for High Throughput Computing (CHTC) at the University of Wisconsin{\textendash}Madison,
Open Science Grid (OSG),
Advanced Cyberinfrastructure Coordination Ecosystem: Services {\&} Support (ACCESS),
Frontera computing project at the Texas Advanced Computing Center,
U.S. Department of Energy-National Energy Research Scientific Computing Center,
Particle astrophysics research computing center at the University of Maryland,
Institute for Cyber-Enabled Research at Michigan State University,
and Astroparticle physics computational facility at Marquette University;
Belgium {\textendash} Funds for Scientific Research (FRS-FNRS and FWO),
FWO Odysseus and Big Science programmes,
and Belgian Federal Science Policy Office (Belspo);
Germany {\textendash} Bundesministerium f{\"u}r Bildung und Forschung (BMBF),
Deutsche Forschungsgemeinschaft (DFG),
Helmholtz Alliance for Astroparticle Physics (HAP),
Initiative and Networking Fund of the Helmholtz Association,
Deutsches Elektronen Synchrotron (DESY),
and High Performance Computing cluster of the RWTH Aachen;
Sweden {\textendash} Swedish Research Council,
Swedish Polar Research Secretariat,
Swedish National Infrastructure for Computing (SNIC),
and Knut and Alice Wallenberg Foundation;
European Union {\textendash} EGI Advanced Computing for research;
Australia {\textendash} Australian Research Council;
Canada {\textendash} Natural Sciences and Engineering Research Council of Canada,
Calcul Qu{\'e}bec, Compute Ontario, Canada Foundation for Innovation, WestGrid, and Compute Canada;
Denmark {\textendash} Villum Fonden, Carlsberg Foundation, and European Commission;
New Zealand {\textendash} Marsden Fund;
Japan {\textendash} Japan Society for Promotion of Science (JSPS)
and Institute for Global Prominent Research (IGPR) of Chiba University;
Korea {\textendash} National Research Foundation of Korea (NRF);
Switzerland {\textendash} Swiss National Science Foundation (SNSF);
United Kingdom {\textendash} Department of Physics, University of Oxford.

\end{document}